\documentclass[lettersize,journal]{IEEEtran}

\usepackage{amsmath,amsfonts}
\usepackage{algorithm}
\usepackage{algorithmic}
\usepackage{array}
\usepackage[caption=false,font=normalsize,labelfont=sf,textfont=sf]{subfig}
\usepackage{textcomp}
\usepackage{stfloats}
\usepackage{url}
\usepackage{verbatim}
\usepackage{graphicx}
\usepackage{cite}
\usepackage{bm}
\usepackage{booktabs}
\usepackage{tabularx}
\usepackage{hyperref}
\usepackage{tikz}
\usepackage{siunitx}
\usepackage{enumitem}

\def\ecrBaseline{46.6}
\def\ecrPUDTune{3.3}
\def\majxThroughputBaseline{0.89}
\def\majxThroughputPUDTune{1.62}
\def\majxSpeedup{1.81}
\def\additionThroughputBaseline{50.2}
\def\additionThroughputPUDTune{94.6}
\def\additionSpeedup{1.88}
\def\multiplicationThroughputBaseline{5.8}
\def\multiplicationThroughputPUDTune{11.0}
\def\multiplicationSpeedup{1.89}
\def\majxThroughputGrainSpeedup{1.03}
\def\majxThroughputWideSpeedup{1.48}
\def\capacityOverhead{0.6}
\def\temperatureReliability{0.14}
\def\timeReliability{0.27}

\def\pudRef{gaoComputeDRAMInMemoryCompute2019,olgunQUACTRNGHighThroughputTrue2021a,gaoFracDRAMFractionalValues2022,kuboMVDRAMEnablingGeMV2025}
\def\pudCite{\cite{\pudRef}}

\newcommand{\MAJ}[1]{\texttt{MAJ#1}}
\newcommand{\RowCopy}{\texttt{RowCopy}}
\newcommand{\Frac}{\texttt{Frac}}
\newcommand{\ACT}{\texttt{ACT}}
\newcommand{\PRE}{\texttt{PRE}}

\graphicspath{{fig/}}  \newcommand*\circled[1]{\tikz[baseline=(char.base)]{\node[circle,draw,inner sep=0.5pt] (char) {#1};}}  

\begin{document}

\title{PUDTune: Multi-Level Charging for High-Precision Calibration in Processing-Using-DRAM}

\author{
    \IEEEauthorblockN{\IEEEauthorrefmark{1}Tatsuya Kubo\IEEEauthorrefmark{2}, \IEEEauthorrefmark{1}Daichi Tokuda\IEEEauthorrefmark{2}, Lei Qu\IEEEauthorrefmark{4}, \IEEEauthorrefmark{1}Ting Cao\IEEEauthorrefmark{4}, and Shinya Takamaeda-Yamazaki\IEEEauthorrefmark{2}\IEEEauthorrefmark{5}}\\
    \IEEEauthorblockA{\IEEEauthorrefmark{2}The University of Tokyo, \IEEEauthorrefmark{4}Microsoft Research, \IEEEauthorrefmark{5}RIKEN}
    \thanks{\IEEEauthorrefmark{1}Corresponding authors: Tatsuya Kubo \texttt{<}tatsuya.kubo@is.s.u-tokyo.ac.jp\texttt{>}, Daichi Tokuda \texttt{<}tokuda-daichi@is.s.u-tokyo.ac.jp\texttt{>}, and Ting Cao \texttt{<}ting.cao@microsoft.com\texttt{>}}
}

\maketitle
\thispagestyle{plain}
\pagestyle{plain}

\begin{abstract}
Recently, practical analog in-memory computing has been realized using unmodified commercial DRAM modules.
The underlying Processing-Using-DRAM (PUD) techniques enable high-throughput bitwise operations directly within DRAM arrays.
However, the presence of inherent error-prone columns hinders PUD's practical adoption.
While selectively using only error-free columns would ensure reliability, this approach significantly reduces PUD's computational throughput.

This paper presents PUDTune, a novel high-precision calibration technique for increasing the number of error-free columns in PUD.
PUDTune compensates for errors by applying pre-identified column-specific offsets to PUD operations.
By leveraging multi-level charge states of DRAM cells, PUDTune generates fine-grained and wide-range offset variations despite the limited available rows.
Our experiments with DDR4 DRAM demonstrate that PUDTune increases the number of error-free columns by \majxSpeedup{}$\times$ compared to conventional implementations, improving addition and multiplication throughput by \additionSpeedup{}$\times$ and \multiplicationSpeedup{}$\times$ respectively.
\end{abstract}

\section{Introduction}
\IEEEPARstart{I}{n} recent years, practical analog in-memory computing has been demonstrated using unmodified commercial DRAM~\pudCite{}.
Through specialized timing control of commands issued by the processor chip, these techniques enable high-throughput computations directly within DRAM arrays.
For instance, MVDRAM~\cite{kuboMVDRAMEnablingGeMV2025} accelerates matrix-vector multiplication for large language model inference within commercial DRAM on real systems.
By leveraging existing memory components for both storage and computation, these approaches offer a practical path to enhance system performance by overcoming the memory bandwidth bottleneck for data-intensive applications.

\begin{figure}[t]
    \centering
    \subfloat[]{
        \includegraphics[width=0.47\columnwidth]{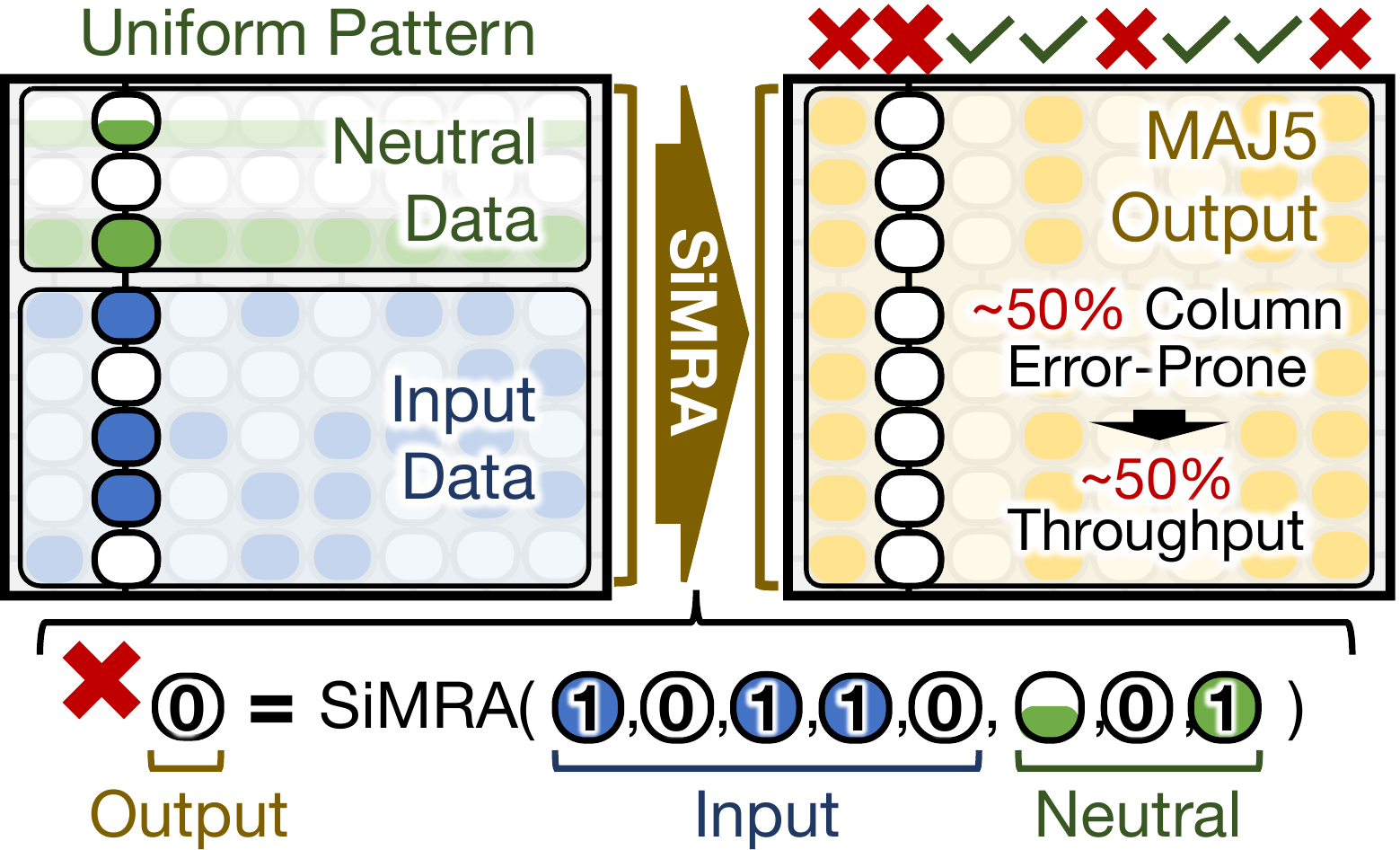}
        \label{fig:01-Introduction1_1}
    }
    \hfill\vrule\hfill
    \subfloat[]{
        \includegraphics[width=0.47\columnwidth]{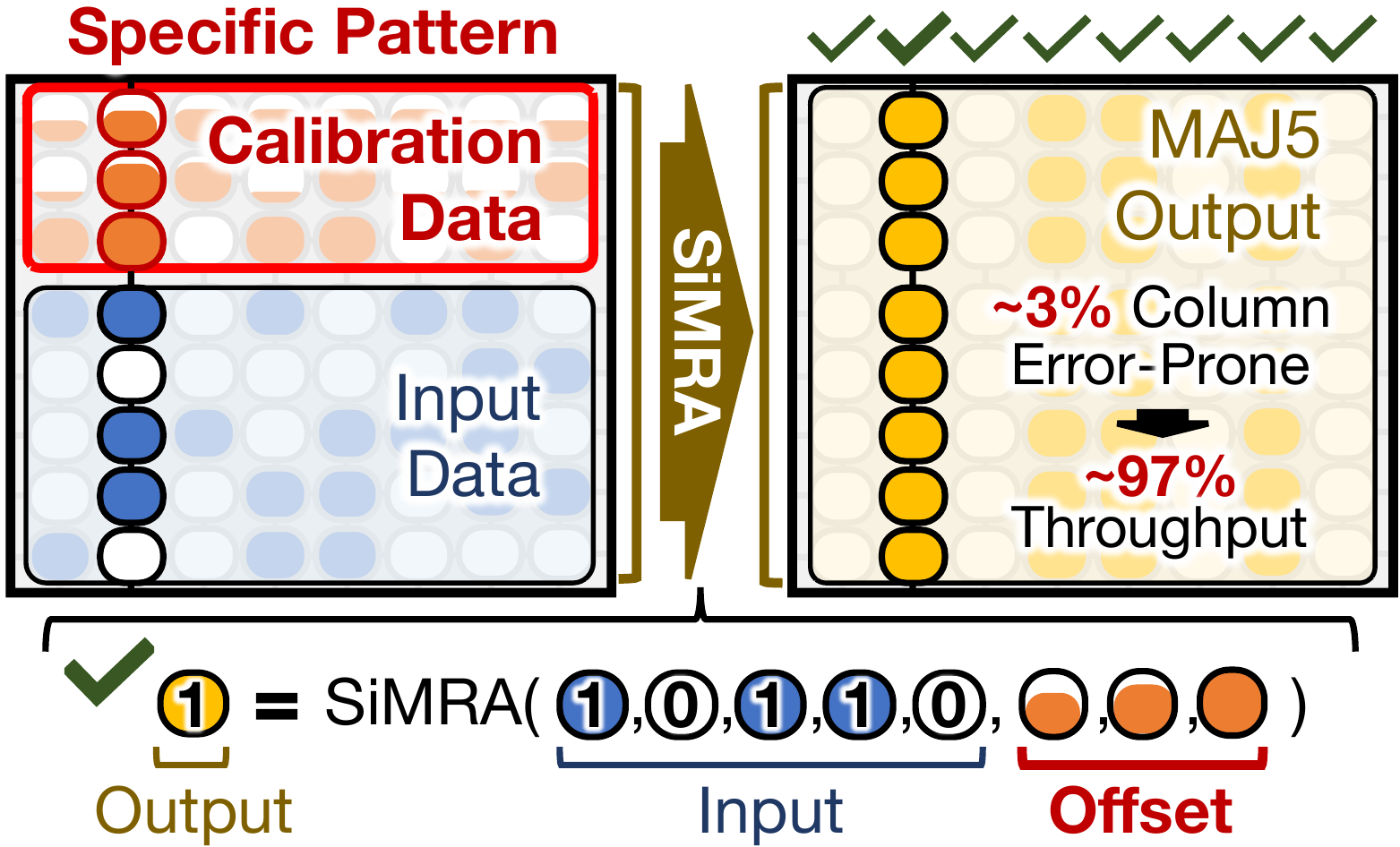}
        \label{fig:01-Introduction1_2}
    }
    \caption{Comparison of \MAJ{5} execution with $8$-row SiMRA in a DRAM subarray. (a) Conventional method. (b) PUDTune.}
    \label{fig:01-Introduction1}
\end{figure}

In the underlying Processing-Using-DRAM (PUD) techniques, \textit{majority-of-X} (\MAJ{X}) operations serve as the fundamental computational primitive~\cite{seshadriAmbitInmemoryAccelerator2017,gaoComputeDRAMInMemoryCompute2019,gaoFracDRAMFractionalValues2022}.
A \MAJ{X} operation determines whether `1' or `0' bits are more prevalent among $X$ bits.
PUD implements this by \textit{simultaneous multi-row activation} (SiMRA), which causes charge sharing across multiple cells~\cite{gaoComputeDRAMInMemoryCompute2019,olgunQUACTRNGHighThroughputTrue2021a,gaoFracDRAMFractionalValues2022}.
Fig.~\ref{fig:01-Introduction1_1} illustrates the conventional implementation of \MAJ{5} using $8$-row SiMRA.
By setting the first three rows to neutral data across all columns, the SiMRA operation results in the \MAJ{5} output for the five inputs.
By constructing majority-based computational graphs, PUD enables primitive operations (\texttt{AND}/\texttt{OR}~\cite{gaoComputeDRAMInMemoryCompute2019} and full-adder~\cite{kuboMVDRAMEnablingGeMV2025}) and complex calculations including matrix multiplication~\cite{kuboMVDRAMEnablingGeMV2025} directly within DRAM.

Despite the potential of PUD, the presence of \textit{error-prone columns} hinders its practical adoption.
The error-prone columns produce erroneous results during \MAJ{X} operations.
These errors arise from threshold voltage variations in sense amplifiers, which, while acceptable for standard DRAM operations, cause certain columns to produce incorrect results during the precise charge sharing process required for \MAJ{X}.
While selectively using only error-free columns would ensure computational reliability, this approach substantially reduces PUD's overall throughput.

In this paper, we present PUDTune, a novel calibration technique for PUD.
As shown in Fig.~\ref{fig:01-Introduction1_2}, PUDTune compensates for errors by applying pre-identified column-specific offsets to counteract the deviations of threshold voltage in sense amplifiers.
Our key insight is that PUD's unique \Frac{}~\cite{gaoFracDRAMFractionalValues2022} can create multi-level charge states within DRAM cells.
By combining the different numbers of \Frac{} operations, PUDTune generates both fine-grained and wide-range offset variations despite the limited available rows.
Our experimental evaluation demonstrates that PUDTune increases the number of error-free columns by \majxSpeedup{}$\times$ compared to conventional implementations, improving the throughput of $8$-bit addition and multiplication operations by \additionSpeedup{}$\times$ and \multiplicationSpeedup{}$\times$ respectively, while requiring only \capacityOverhead{}\% capacity overhead. \section{Background and Motivation}
\subsection{Dynamic Random Access Memory}
\begin{figure}[t]
    \centering
    \subfloat[]{
        \includegraphics[height=0.29\columnwidth,keepaspectratio]{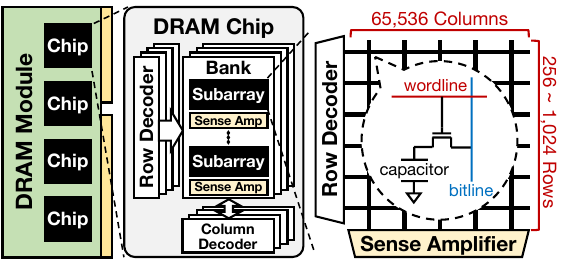}
        \label{fig:02-Background1_1}
    }
    \subfloat[]{
        \includegraphics[height=0.29\columnwidth,keepaspectratio]{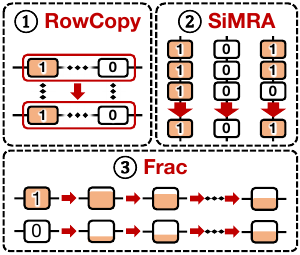}
        \label{fig:02-Background1_2}
    }
    \caption{(a) DRAM organization. (b) PUD operations.}
    \label{fig:02-Background1}
\end{figure}

DRAM serves as the primary memory in modern computing systems, organized in a hierarchical structure as shown in Fig.~\ref{fig:02-Background1_1}.
A DRAM system consists of multiple channels, each containing DRAM chips divided into banks.
Each bank contains multiple subarrays, with each subarray comprising 256-1,024 rows and 65,536 columns of memory cells.
Individual memory cells store a single bit and connect to a wordline (row) and bitline (column).

The memory controller orchestrates DRAM commands: \ACT{} (activation) opens a row and copies data to the row buffer, while \PRE{} (precharge) closes the active row and prepares for the next activation.
These commands adhere to specific timing constraints to maintain data integrity during normal operations.

\subsection{Processing-Using-DRAM}\label{sec:02-PUD}
Processing-Using-DRAM (PUD) exploits the analog operational characteristics of DRAM to perform highly parallel bit-serial computations directly within memory arrays~\cite{seshadriAmbitInmemoryAccelerator2017}.
Commercial off-the-shelf DRAM can support PUD without hardware modifications by intentionally violating timing parameters specified by manufacturers~\pudCite{}.

PUD with unmodified DRAM supports three fundamental operations (Fig.~\ref{fig:02-Background1_2}):
(1) \RowCopy{}~\cite{gaoComputeDRAMInMemoryCompute2019} transfers data between rows within the same subarray.
(2) \textit{Simultaneous many-row activation} (SiMRA)~\cite{gaoComputeDRAMInMemoryCompute2019,olgunQUACTRNGHighThroughputTrue2021a} activates multiple rows simultaneously, allowing their charge to be shared across the same column.
(3) \Frac{}~\cite{gaoFracDRAMFractionalValues2022} applies partial charges to cells, creating intermediate states between `0' and `1'.

The \textit{majority-of-X} (\MAJ{X}) operation, implemented through a combination of these primitives, performs a majority vote among $X$ cells of the same column.
For instance, \MAJ{5} implementation typically follows a specific execution flow (Fig.~\ref{fig:01-Introduction1_1}):
\circled{1} 5 input rows and 3 neutral rows are arranged in designated rows by \RowCopy{};
\circled{2} the first neutral row is charged by \Frac{} to a half-charged state;
\circled{3} 8 rows are simultaneously activated through SiMRA;
\circled{4} the majority result is stored in all the 8 rows.

\subsection{Error-Prone Columns}

Despite the potential of PUD, the presence of \textit{error-prone columns} hinders the practical adoption of PUD.
The error-prone columns produce erroneous results during \MAJ{X} operations.
These errors arise primarily from the threshold voltage variation in sense amplifiers~\cite{kimModelBasedVariationAwareOptimization2025}.
A sense amplifier detects small voltage differences between the bitline and its threshold, amplifying these differences to recognizable logic levels (e.g., 0 or 1).
Ideally, the threshold should be set at $0.5V_{DD}$, but due to process variation, this threshold can deviate slightly, for example, to $0.48V_{DD}$ or $0.53V_{DD}$.

PUD operations require higher precision in threshold voltage than standard DRAM operations.
For example, in a DRAM read, a cell capacitor of $30 \si{\femto\farad}$ and a bitline of $270 \si{\femto\farad}$ would result in a voltage of $0.55V_{DD}$ after charge sharing.
This is distinguishable from the threshold even with a 5\% deviation.
However, the charge sharing among multiple cells in PUD operations reduces the voltage difference.
For \MAJ{5}(1, 1, 1, 0, 0) with 8-row SiMRA, which should output 1, the voltage will be around $0.529V_{DD}$, where the sense amplifier with 5\% deviation might incorrectly output 0.

Selective use of error-free columns ensures computational reliability but degrades throughput in PUD operations.
The throughput of \MAJ{X} operations can be expressed as:
\begin{align}
    \text{Throughput} = \frac{\text{\# of Error-Free Columns}}{\text{Latency of \MAJ{X}}},
    \label{eq:02-Error}
\end{align}
where error-free columns produce consistent results during \MAJ{X} operations.
In particular, the reliability of \MAJ{5} operations bottlenecks the performance.
Since PUD arithmetic operations using full-adders utilize \MAJ{5}~\cite{kuboMVDRAMEnablingGeMV2025}, the overall system throughput is limited by \MAJ{5} reliability.
Our preliminary experiments show that \MAJ{5} operations degrade to approximately 50\% with SK Hynix DDR4 DRAM modules.
 \begin{figure*}[t]
    \centering
    \subfloat[]{
        \includegraphics[width=0.31\textwidth]{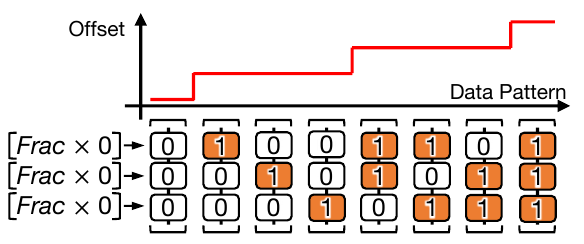}
        \label{fig:03-PUDTune3_1}
    }
    \hfill
    \subfloat[]{
        \includegraphics[width=0.31\textwidth]{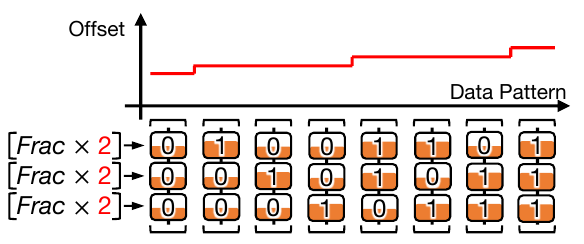}
        \label{fig:03-PUDTune3_2}
    }
    \hfill
    \subfloat[]{
        \includegraphics[width=0.31\textwidth]{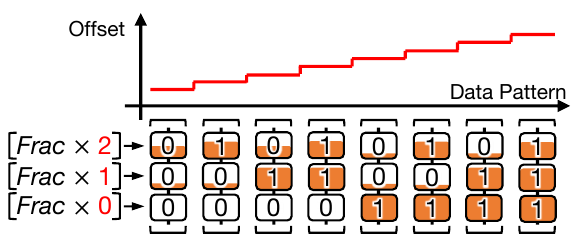}
        \label{fig:03-PUDTune3_3}
    }
    \caption{Comparison of offset variations for different \Frac{} count configurations. (a) $\text{T}_{0, 0, 0}$: No \Frac{} operations applied. (b) $\text{T}_{2, 2, 2}$: Two \Frac{} operations applied to all three rows. (c) $\text{T}_{2, 1, 0}$: Two, one, and zero \Frac{} operations applied to the first, second, and third rows, respectively.}
    \label{fig:03-PUDTune3}
\end{figure*}
\section{PUDTune}
\subsection{Goal}
The goal of this paper is to reduce the proportion of error-prone columns and increase operational throughput by introducing \textit{offset calibration}~\cite{kimModelBasedVariationAwareOptimization2025,yu65nm8TSRAM2022} to PUD.
The offset calibration technique, especially common in SRAM-based approaches~\cite{yu65nm8TSRAM2022}, allocates specific cells in the dedicated rows to counteract the column-specific offsets caused by process variations.
Our approach replaces the uniform neutral data in PUD with column-specific calibration data (Fig.~\ref{fig:01-Introduction1_2}).

\textbf{Calibration data identification is required only once per manufactured DRAM device}.
We identify calibration data for each subarray in the manufactured DRAM, then repeatedly apply this calibration data during each \MAJ{X} execution.
By storing the bit patterns used for calibration data generation in non-volatile memory, it can be reused across different environments and system reboots.

In the following sections, we primarily focus on the case of \MAJ{5}.
The \MAJ{5} operation, which is utilized in full-adder~\cite{kuboMVDRAMEnablingGeMV2025}, exhibits low accuracy and is a throughput bottleneck.
Note that PUDTune can be naturally extended to \MAJ{X} operations with different input sizes.

\subsection{Challenge}
The challenge in applying offset calibration to PUD is the limited number of available rows.
The small number of calibration rows restricts the variety of offsets, making it difficult to comprehensively adapt to the distribution of threshold voltage variations.
In a SRAM-based approach that employs offset calibration, 32 out of 256 simultaneously opened rows are allocated for calibration purposes~\cite{yu65nm8TSRAM2022}.
This allows for the adjustment of convergence voltage offsets in 33 levels with a granularity of 1/256 of the supply voltage.
In contrast, \MAJ{5} operations implemented with $8$-row SiMRA can only repurpose $3$ non-operand rows for calibration.
This limits offset adjustments to just $4$ levels with a coarse granularity of $1/8$ of the supply voltage.

\subsection{Insight: Multi-Level Charging}
The key observation is the multi-level charge states via \Frac{} operations~\cite{gaoFracDRAMFractionalValues2022}, which is unique to PUD.
When \Frac{} operations are repeatedly applied to the same row, the cell values gradually approach a neutral charge state (Fig.~\ref{fig:02-Background1_2}).
Prior works typically use 6-10 \Frac{} operations to achieve this neutral state~\cite{gaoFracDRAMFractionalValues2022}.
We observe that applying fewer \Frac{} operations results in intermediate charge states between the initial value and the neutral state.
This property of multi-level charge states has been experimentally validated through retention time measurements~\cite{gaoFracDRAMFractionalValues2022}.

\subsection{Proposal: PUDTune}
To address the challenge of limited rows, we propose PUDTune, a novel high-precision offset calibration method for PUD.
The key idea of PUDTune is to leverage the combination of different-level charge states to accommodate a finer grain and wider range of offset variations with the limited available rows.
Through this improved offset diversity, PUDTune can better adapt to the distribution of threshold voltage variations across columns, enabling compensation for a larger proportion of error-prone columns.
Fig.~\ref{fig:03-PUDTune3} illustrates the relationship between calibration data patterns and the variety of offsets generated for different \Frac{} count configurations.
Without \Frac{} operations ($\text{T}_{0,0,0}$), offsets cover a wide range but with coarse granularity.
With uniform \Frac{} operations ($\text{T}_{2,2,2}$), offsets have finer granularity but narrower range.
By applying different numbers of \Frac{} operations to each row ($\text{T}_{2,1,0}$), PUDTune achieves both fine-grain and wide-range offset variations.

\subsubsection*{Method}
PUDTune's \MAJ{X} execution modifies the execution flow described in Sec.~\ref{sec:02-PUD} as follows:
\circled{1}' We first move pre-identified calibration data, which are reserved within the subarray, along with input data to rows opened by SiMRA using \RowCopy{} operations.
\circled{2}' We then apply the configured number of \Frac{} operations to each calibration row.
The execution latency varies based on the total \Frac{} operations used, potentially achieving lower latency than conventional methods when fewer operations are applied.
While PUDTune requires storing calibration data within the subarray, the implementation for \MAJ{3} and \MAJ{5} operations necessitates only three rows, resulting in \capacityOverhead{}\% capacity overhead.

\subsubsection*{Calibration Data Identification}
\begin{algorithm}[t]
    \caption{Calibration Data Identification}
    \label{alg:03-PUDTune}
    \begin{algorithmic}[1]
        \STATE \textbf{Input/Output:} calibration\_data
        \FOR{iteration = 1 to n\_iterations}
            \STATE store\_to\_dram(calibration\_data)
            \STATE results = majx\_sampling()
            \FOR{i = 1 to n\_columns}
                \STATE bias = get\_bias(results[i])
                \IF{bias \textgreater{} threshold}
                    \STATE decrement\_level(calibration\_data[i])
                \ELSIF{bias \textless{} -threshold}
                    \STATE increment\_level(calibration\_data[i])
                \ENDIF
            \ENDFOR
        \ENDFOR
    \end{algorithmic}
\end{algorithm}
We identify calibration data through an iterative algorithm (Algorithm~\ref{alg:03-PUDTune}).
For each iteration, we record \MAJ{X} outputs for each column across random input patterns.
We calculate a \textit{bias} metric representing the proportion of '1' outputs for each column.
When this bias exceeds a predefined threshold, we update the corresponding calibration data to counteract the offset causing the bias.
This process iteratively refines the calibration data, identifying the optimal patterns to minimize errors across all columns. \section{Evaluation}
\begin{figure}[t]
    \centering
    \includegraphics[width=0.8\columnwidth]{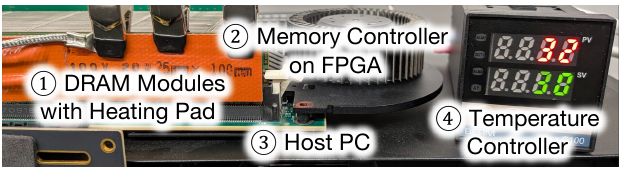}
    \caption{Experimental setup.}
    \label{fig:04-Evaluation1}
\end{figure}
\subsection{Experimental Methodology}
Fig.~\ref{fig:04-Evaluation1} shows our experimental setup consisting of four key components:
\circled{1} DRAM modules equipped with heating pads.
We use 48 DDR4-2133 chips selected from 16 SK Hynix modules, which are capable of SiMRA operations.
\circled{2} A memory controller on FPGA.
We utilize DRAM Bender~\cite{olgunDRAMBenderExtensible2023}, an open-source DRAM controller, which is implemented on a Xilinx Alveo U200 for precise DRAM manipulation.
\circled{3} A host PC that controls the FPGA.
\circled{4} A temperature controller for the heating pads to evaluate thermal reliability.

In our evaluation, we compare two methods of \MAJ{5} implementation parameterized by the number of \Frac{} operations:
\begin{itemize}[leftmargin=*]
    \item \textbf{$\text{B}_{x, 0, 0}$ (Baseline)}:
        The baseline \MAJ{5} implementation, where we apply $x$ times of \Frac{} operations to the first of the three non-operand rows, while storing constants 0 and 1 in the remaining two rows.
    \item \textbf{$\text{T}_{x, y, z}$ (PUDTune)}:
        The PUDTune \MAJ{5} implementation, where we apply pre-identified calibration data to all three non-operand rows with $x$, $y$, and $z$ times of \Frac{} operations applied to each row respectively.
\end{itemize}

Our experiments evaluate two key performance factors:
\begin{itemize}[leftmargin=*]
    \item \textbf{Error-prone column ratio (ECR)}:
        We define ECR as the percentage of columns that output no errors across all 65,536 rows in a subarray.
        We measure ECR by testing with 8,192 random inputs for each bank in the tested DRAM modules.
\item \textbf{Throughput}:
        We calculate throughput for a system with 4-channel DRAM using Eq.~\ref{eq:02-Error}.
        The latency is derived from the 16 bank-parallel PUD under \ACT{} power constraints.
\end{itemize}

We identify calibration data using 20 iterations of our algorithm, collecting 512 random samples per iteration.
The entire calibration process takes approximately 1 minute per subarray using our Python implementation on DRAM Bender.

\subsection{Results}
\begin{table}[h]
    \caption{ECR and throughput.}
    \centering
    \label{tab:04-Evaluation2}
    \begin{tabular}{ccccc}
        \toprule
        \textbf{Method} & \textbf{ECR} & \textbf{MAJ5} & \textbf{$8$-bit ADD} & \textbf{$8$-bit MUL} \\
        \midrule
        Baseline ($\text{B}_{3, 0, 0}$) & \ecrBaseline{}\% & \majxThroughputBaseline{} TOPS & \additionThroughputBaseline{} GOPS & \multiplicationThroughputBaseline{} GOPS \\
        PUDTune ($\text{T}_{2, 1, 0}$) & \ecrPUDTune{}\% & \majxThroughputPUDTune{} TOPS & \additionThroughputPUDTune{} GOPS & \multiplicationThroughputPUDTune{} GOPS \\
        \bottomrule
    \end{tabular}
\end{table}
\subsubsection{ECR and throughput.}
Table~\ref{tab:04-Evaluation2} compares the error-prone column ratio (ECR) and computational throughput between the baseline and PUDTune implementations.
We can observe that PUDTune reduces ECR from \ecrBaseline{}\% in the baseline to \ecrPUDTune{}\% on average.
This reduction in error-prone columns increases the number of error-free columns available for computation, providing \majxSpeedup{}$\times$ improvement in \MAJ{5} throughput.
Additionally, it can be seen that PUDTune achieves \additionSpeedup{}$\times$ and \multiplicationSpeedup{}$\times$ improvement in throughput for $8$-bit addition and multiplication compared to the baseline implementation.

\begin{figure}[h]
    \centering
    \includegraphics[width=\columnwidth]{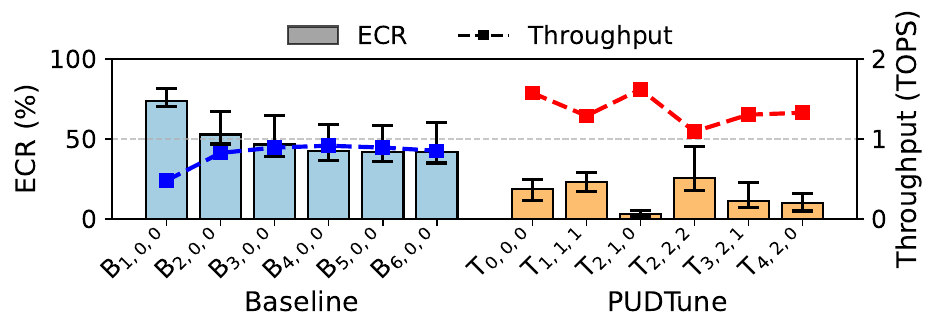}
    \caption{\MAJ{5} performance sensitivity to \Frac{} times.}
    \label{fig:04-Evaluation2}
\end{figure}
\subsubsection{Sensitivity to \Frac{} times.}
Fig.~\ref{fig:04-Evaluation2} compares the performance sensitivity of \MAJ{5} to the number of \Frac{} operations.
PUDTune consistently outperforms the baseline with lower ECR and higher throughput across all configurations.
The $\text{T}_{2,1,0}$ configuration delivers optimal results, achieving \majxThroughputGrainSpeedup{}$\times$ improvement over $\text{T}_{0,0,0}$ and \majxThroughputWideSpeedup{}$\times$ over $\text{T}_{2,2,2}$.
These results confirm that $\text{T}_{2,1,0}$ effectively balances fine-grained precision with wide-range offset variations, maximizing available error-free columns.

\begin{figure}[h]
    \centering
    \subfloat[]{
        \includegraphics[width=0.48\columnwidth]{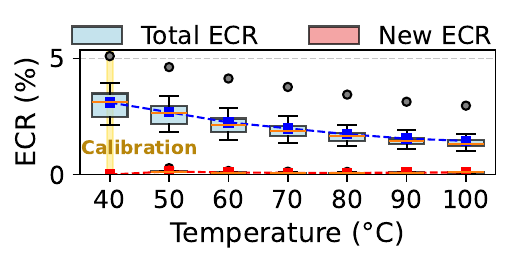}
        \label{fig:04-Evaluation3_1}
    }
\subfloat[]{
        \includegraphics[width=0.48\columnwidth]{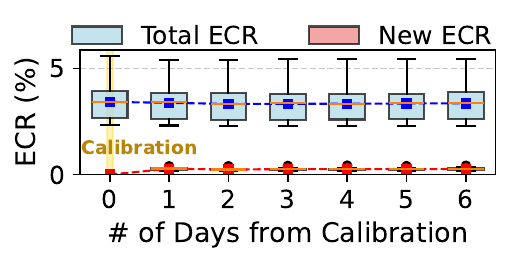}
        \label{fig:04-Evaluation3_2}
    }
    \caption{Reliability analysis. (a) Temperature. (b) Time.}
    \label{fig:04-Evaluation3}
\end{figure}
\subsubsection{Reliability Analysis}
We evaluate PUDTune's reliability by measuring ECR and new error-prone columns under various conditions.
Using $\text{T}_{2,1,0}$ configuration, we conducted temperature tests from 40$^{\circ}$C to 100$^{\circ}$C and time-based evaluations over one week.
Fig.~\ref{fig:04-Evaluation3_1} shows that new ECR remains below \temperatureReliability{}\% from 40$^{\circ}$C to 100$^{\circ}$C, demonstrating PUDTune's thermal reliability.
Fig.~\ref{fig:04-Evaluation3_2} shows new ECR stays below \timeReliability{}\% throughout the week, confirming PUDTune's long-term reliability.
 \section{Conclusion}
We present PUDTune, a novel high-precision calibration technique for PUD, achieving \majxSpeedup{}$\times$ improvement in the number of error-free columns.
PUDTune maintains compatibility with existing DRAM manufacturing methods, making it an essential technique for practical PUD applications.

\section*{Acknowledgements}
This work is supported in part by JSPS KAKENHI Grant Number 23H00467 and JST CREST JPMJCR21D2.

\bibliographystyle{IEEEtran}

\begin{thebibliography}{1}
\providecommand{\url}[1]{#1}
\csname url@samestyle\endcsname
\providecommand{\newblock}{\relax}
\providecommand{\bibinfo}[2]{#2}
\providecommand{\BIBentrySTDinterwordspacing}{\spaceskip=0pt\relax}
\providecommand{\BIBentryALTinterwordstretchfactor}{4}
\providecommand{\BIBentryALTinterwordspacing}{\spaceskip=\fontdimen2\font plus
\BIBentryALTinterwordstretchfactor\fontdimen3\font minus
  \fontdimen4\font\relax}
\providecommand{\BIBforeignlanguage}[2]{{\expandafter\ifx\csname l@#1\endcsname\relax
\typeout{** WARNING: IEEEtran.bst: No hyphenation pattern has been}\typeout{** loaded for the language `#1'. Using the pattern for}\typeout{** the default language instead.}\else
\language=\csname l@#1\endcsname
\fi
#2}}
\providecommand{\BIBdecl}{\relax}
\BIBdecl

\bibitem{gaoComputeDRAMInMemoryCompute2019}
F.~Gao, G.~Tziantzioulis, and D.~Wentzlaff, ``{{ComputeDRAM}}: {{In-Memory
  Compute Using Off-the-Shelf DRAMs}},'' in \emph{Proceedings of the 52nd
  {{Annual IEEE}}/{{ACM International Symposium}} on
  {{Microarchitecture}}}.\hskip 1em plus 0.5em minus 0.4em\relax Columbus OH
  USA: ACM, Oct. 2019, pp. 100--113.

\bibitem{olgunQUACTRNGHighThroughputTrue2021a}
A.~Olgun, M.~Patel, A.~G. Ya{\u g}l{\i}k{\c c}{\i}, H.~Luo, J.~S. Kim,
  F.~Nisa~Bostanc{\i}, N.~Vijaykumar, O.~Ergin, and O.~Mutlu, ``{{QUAC-TRNG}}:
  {{High-Throughput True Random Number Generation Using Quadruple Row
  Activation}} in {{Commodity DRAM Chips}},'' in \emph{2021 {{ACM}}/{{IEEE}}
  48th {{Annual International Symposium}} on {{Computer Architecture}}
  ({{ISCA}})}, Jun. 2021, pp. 944--957.

\bibitem{gaoFracDRAMFractionalValues2022}
F.~Gao, G.~Tziantzioulis, and D.~Wentzlaff, ``{{FracDRAM}}: {{Fractional
  Values}} in {{Off-the-Shelf DRAM}},'' in \emph{2022 55th {{IEEE}}/{{ACM
  International Symposium}} on {{Microarchitecture}} ({{MICRO}})}.\hskip 1em
  plus 0.5em minus 0.4em\relax Chicago, IL, USA: IEEE, Oct. 2022, pp. 885--899.

\bibitem{kuboMVDRAMEnablingGeMV2025}
T.~Kubo, D.~Tokuda, T.~Nagatani, M.~Usui, L.~Qu, T.~Cao, and
  S.~{Takamaeda-Yamazaki}, ``{{MVDRAM}}: {{Enabling GeMV Execution}} in
  {{Unmodified DRAM}} for {{Low-Bit LLM Acceleration}},'' Mar. 2025.

\bibitem{seshadriAmbitInmemoryAccelerator2017}
V.~Seshadri, D.~Lee, T.~Mullins, H.~Hassan, A.~Boroumand, J.~Kim, M.~A. Kozuch,
  O.~Mutlu, P.~B. Gibbons, and T.~C. Mowry, ``Ambit: In-memory accelerator for
  bulk bitwise operations using commodity {{DRAM}} technology,'' in
  \emph{Proceedings of the 50th {{Annual IEEE}}/{{ACM International Symposium}}
  on {{Microarchitecture}}}.\hskip 1em plus 0.5em minus 0.4em\relax Cambridge
  Massachusetts: ACM, Oct. 2017, pp. 273--287.

\bibitem{kimModelBasedVariationAwareOptimization2025}
D.~Kim, G.~Kim, S.~Kim, J.~Park, S.~Kim, H.~Seo, C.~Lim, S.~Kim, J.~Lee,
  J.~Yun, H.~Lee, J.~Park, Y.~Lee, S.~Lee, and M.~Lee, ``Model-{{Based
  Variation-Aware Optimization}} for {{Offset Calibration}} and {{Pre-Sensing}}
  in {{DRAM Sense Amplifiers}},'' \emph{IEEE Access}, pp. 1--1, 2025.

\bibitem{yu65nm8TSRAM2022}
C.~Yu, T.~Yoo, K.~T.~C. Chai, T.~T.-H. Kim, and B.~Kim, ``A 65-nm {{8T SRAM
  Compute-in-Memory Macro With Column ADCs}} for {{Processing Neural
  Networks}},'' \emph{IEEE Journal of Solid-State Circuits}, vol.~57, no.~11,
  pp. 3466--3476, Nov. 2022.

\bibitem{olgunDRAMBenderExtensible2023}
A.~Olgun, H.~Hassan, A.~G. Ya{\u g}l{\i}k{\c c}{\i}, Y.~C. Tu{\u g}rul,
  L.~Orosa, H.~Luo, M.~Patel, O.~Ergin, and O.~Mutlu, ``{{DRAM Bender}}: {{An
  Extensible}} and {{Versatile FPGA-based Infrastructure}} to {{Easily Test
  State-of-the-art DRAM Chips}},'' Sep. 2023.

\end{thebibliography}

\end{document}